\documentclass[final]{svjour2}
\usepackage{graphicx}
\usepackage{amssymb}
\usepackage{mathptmx}
\usepackage[numbers]{natbib}
\makeatletter
\journalname{Journal of Low Temperature Physics}

\bibpunct{}{}{,}{s}{}{,}

\begin{document}

\newcommand{\hdblarrow}{H\makebox[0.9ex][l]{$\downdownarrows$}-}
\title{Creation of NOON states by double Fock-state/Bose-Einstein condensates}

\author{W. J. Mullin$^1$ \and  F. Lalo\"{e}$^2$}

\institute{1:Department of Physics, University of Massachusetts, Amherst,
Massachusetts 01003 USA\\
\email{mullin@physics.umass.edu}
\\2: Laboratoire Kastler Brossel, ENS, UPMC, CNRS ; 24 rue Lhomond,
75005 Paris, France\\
\email{laloe@lkb.ens.fr}}

\date{06.02.2010}

\maketitle


\begin{abstract}
NOON states (states of the form $|N>_{a}|0>_{b}+|0>_{a}|N>_{b}$ where
$a$ and $b$ are single particle states) have been used for predicting
violations of hidden-variable theories
(Greenberger-Horne-Zeilinger violations) and are valuable in
metrology for precision measurements of phase at the Heisenberg limit.
We show theoretically how the use of two Fock state/Bose-Einstein
condensates as sources in a modified Mach Zender interferometer can
lead to the creation of the NOON state in which $a$ and $b$ refer to
arms of the interferometer and $N$ is the total number of particles in
the two condensates.  The modification of the interferometer involves
making conditional ``side'' measurements of a few
particles near the sources.  These measurements put the remaining
particles in a superposition of two phase states, which are converted
into NOON states by a beam splitter.  The result is equivalent to the
quantum experiment in which a large molecule passes through two slits.
The NOON states are combined in a final beam splitter and show
interference.  Attempts to detect through which ``slit'' the
condensates passed destroys the interference.

\end{abstract}

\section{Introduction}

NOON states are interesting and useful; they are {}``all-or-nothing''
states, having the form\begin{equation} \left|\Phi\right\rangle
=|N>_{a}|0>_{b}+|0>_{a}|N>_{b}\end{equation} where the subscripts $a$
and $b$ represent single particle states.  The superposition is of all
$N$ particles in state $a$ and none in $b$, plus none in $b$ and all
in $a.$ Such states have had several uses in the past: A) They are the
ultimate Schrodinger cat states, sometimes called ``maximally
entangled''.  One may use them to demonstrate the quantum interference
of macroscopically distinct objects.  \cite{Leggett-1} B) They have
been used to study violations of quantum realism in the well-known
Greenberger-Horne-Zeilinger contradictions.  \cite{GHZ,Mermin} C) They
can be used to violate the standard quantum limit and approach the
Heisenberg limit in metrology.  \cite{metrology, metrology2} D) They
may provide for the possibility of quantum lithography.  \cite{Litho}

Several methods have been proposed to create NOON states by projective
measurement techniques.\cite{Kok, Fiur} The method due to Cable and Dowling \cite{Cable} is quite similar to the one presented here. A two-body NOON state can be
constructed by allowing two bosons to impinge one on either side of a
50-50 beam splitter.  The final state will be a superposition of
two-particles on either side of the splitter according to the
Hong-Ou-Mandel effect.  \cite{HOM} Here we generalize this situation
with an arbitrary number of particles in the sources and show that an
appropriate preparation procedure, using state-vector reduction and a
conditional preparation, leads to the creation of a NOON state.  We
will use two Bose-Einstein condensate/Fock states as sources for an
interferometer.  These number states are assumed to consist entirely
of ground-state bosons, such as those available approximately in
ultra-cold gas systems.  The NOON state is constructed here with $a$
and $b$ representing two arms of the interferometer.  The two
components can be brought together at a beam splitter to cause the
quantum interference.  In this regard the system is rather like having
a large molecule traveling in a superposition through two slits before
interfering with itself on a screen.  In analogy with two-slit
problem, if an observer tries to tell through which arm the N
particles has passed, the interference pattern is ruined.

\section{Interferometer}

\begin{figure}[h]
\centering
\includegraphics[width=4in]{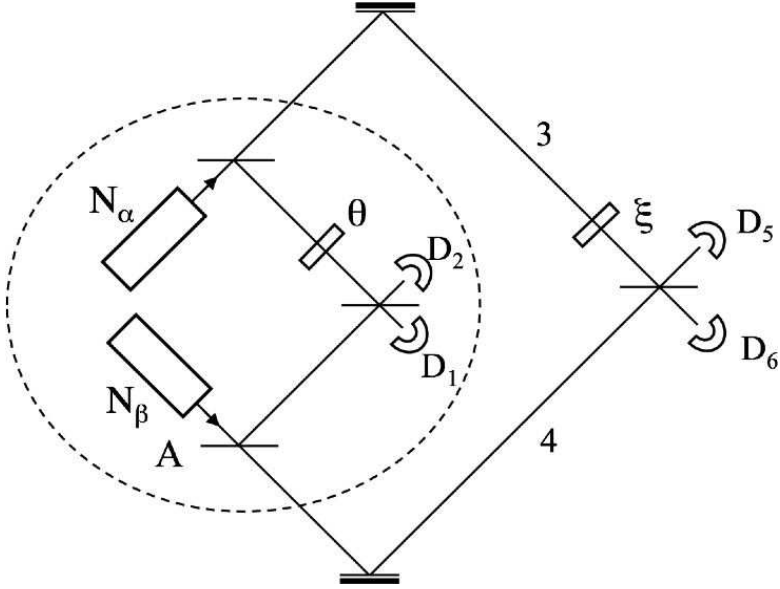}

\caption{Two sources have populations
$N_{\alpha}$ and $N_{\beta}$ (double Fock state).  Some of the emitted
particles are used for a measurement of the relative phase, with the
help of two beam splitters near the sources and an interferometer
with phase shift  $\theta=\pi/2$ and detectors D1 and D2, recording 
$m_{1}$ and $m_{2}$ particles.  When events where $m_{1}=m_{2}$ are selected, the quantum
projection creates a coherent superposition of states  in arms 3 and 4 
where the two beams have opposite relative phases.  When the phase
shift $\xi$ is set to zero, after the last beam splitter this
superposition becomes a NOON states in arms 5 and 6.}

\label{fig1}
\end{figure}

The interferometer is shown in Fig.  \ref{fig1}.  Two Fock state
sources of number $N_{\alpha}$ and $N_{\beta}$ enter the
interferometer.  The side detectors 1 and 2, situated immediately
after the sources, are a key element; by measuring $m_{1}$ and $m_{2}$
particles in these detectors, the remaining particles are put into
phase states that then appear in arms 3 and 4.  When these pass
through the middle beam splitter the result (for suitable value of
$\xi$) is a NOON state in arms 5 and 6.  The nature of the states in
these arms could be tested by examining them in detectors 5 and 6.

The destruction operators at the detectors are found by tracing back
from from a detector to each source. We have

\begin{eqnarray}\label{aops}
a_{1} & = & \frac{1}{2}\left(ie^{i\theta}a_{\alpha}-a_{\beta}\right)\quad\quad a_{2}=\frac{1}{2}\left(-e^{i\theta}a_{\alpha}+ia_{\beta}\right)\nonumber \\
a_{5} & = & \frac{1}{2}\left(-ie^{i\xi}a_{\alpha}-a_{\beta}\right)\quad\quad a_{6}=\frac{1}{2}\left(-e^{i\xi}a_{\alpha}-ia_{\beta}\right)\nonumber \\
\end{eqnarray}

We take $m_{1}$ and $m_{2}$ particles to be deflected by beam splitters
into detectors 1 and 2 respectively. Subsequently if we are looking
at detections in arms $i$ and $j$ then the amplitude for finding
particle numbers $\{m_{1},m_{2},m_{i},m_{j}\}$ is

\begin{equation}
C_{m_{1},m_{2},m_{i},m_{j}}=\left\langle 0\left|\frac{a_{i}^{m_{i}}a_{j}^{m_{j}}a_{1}^{m_{1}}a_{2}^{m_{2}}}{\sqrt{m_{1}!m_{2}!m_{i}!m_{j}!}}\right|N_{\alpha}N_{\beta}\right\rangle \label{C}\end{equation}
 The double Fock state (DFS) $\left|N_{\alpha}N_{\beta}\right\rangle $
can be expanded in phase states as \begin{equation}
\left|N_{\alpha}N_{\beta}\right\rangle =\sqrt{\frac{2^{N}N_{\alpha}!N_{\beta}!}{N!}}\int_{-\pi}^{\pi}\frac{d\phi}{2\pi}e^{-iN_{\beta}\phi}\left|\phi,N\right\rangle \label{eq:PhaseExp}\end{equation}
These states have the property that for any $a_{i}=v_{i\alpha}a_{\alpha}+v_{i\beta}a_{\beta}$:
\begin{equation}
a_{i}\left|\phi,N\right\rangle =\sqrt{\frac{N}{2}}(v_{i\alpha}+v_{i\beta}e^{i\phi})\left|\phi,N-1\right\rangle \end{equation}
so that the state created by the interferometer measurements at 1 and 
2 is \begin{equation}
\left|\Gamma\right\rangle \equiv a_{1}^{m_{1}}a_{2}^{m_{2}}\left|N_{\alpha}N_{\beta}\right\rangle \sim\int_{-\pi}^{\pi}\frac{d\phi}{2\pi}e^{-iN_{\beta}\phi}R_{12}(\phi)\left|\phi,N-M\right\rangle \label{eq:gamma}\end{equation}
where $M=m_{1}+m_{2}$ and \begin{eqnarray}
R_{12}(\phi) & = & (1+e^{i\phi})^{m_{1}}(1-e^{i\phi})^{m_{2}}\label{eq:R12}\end{eqnarray}
We have taken $\theta=\pi/2$, which places the peaks symmetrically
about $\phi=0$. We have \begin{equation}
R_{12}(\phi)=(-i)^{m_{2}}2^{M}e^{iM\phi/2}Q_{12}(\phi)\end{equation}
where\begin{equation}
Q_{12}(\phi)=\left(\cos\frac{\phi}{2}\right)^{m_{1}}\left(\sin\frac{\phi}{2}\right)^{m_{2}}\label{eq:Qphi}\end{equation}

\begin{figure}[h]
\centering \includegraphics[width=4in]{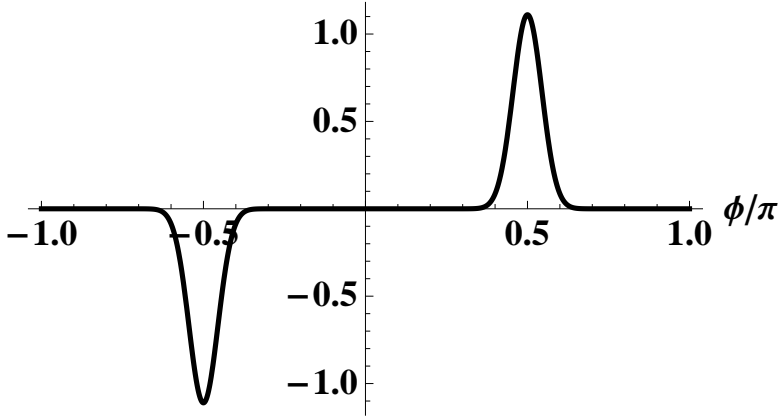}

\caption{Plot of $Q_{12}(\phi)$ of Eq. (\ref{eq:Qphi}) for $m_{1}=m_{2}=52$.
Because we have taken $\theta=\pi/2$ and $m_{1}=m_{2}$ we find the
two symmetrical peaks at $\pm\pi/2.$ For odd $m_{2}$ we have one positive and
one negative peak. With $m_{2}$ even both would be positive.}

\label{figQ}
\end{figure}

For arbitrary $m_{1},m_{2}$ Q$_{12}$ has peaks at $\pm\phi_{0}=\pm2\arctan(m_{2}/m_{1}).$
It will be convenient for us to choose the ensemble of experiments
in which we find $m_{1}=m_{2}$ particles in the initial two branches. In
that case we find the plot of $Q_{12}$ has peaks at $\pm\pi/2$ as
shown in Fig. \ref{figQ}.%
We see that we already have a superposition of phase states in arms
3 and 4 (but not localized in any particular arm). We have
previously used these states to demonstrate macroscopic quantum interference.
\cite{MF PRL} However, now we let them pass through a middle beam
splitter rather than entering detectors. 

\begin{figure}[h]
\centering \includegraphics[width=4in]{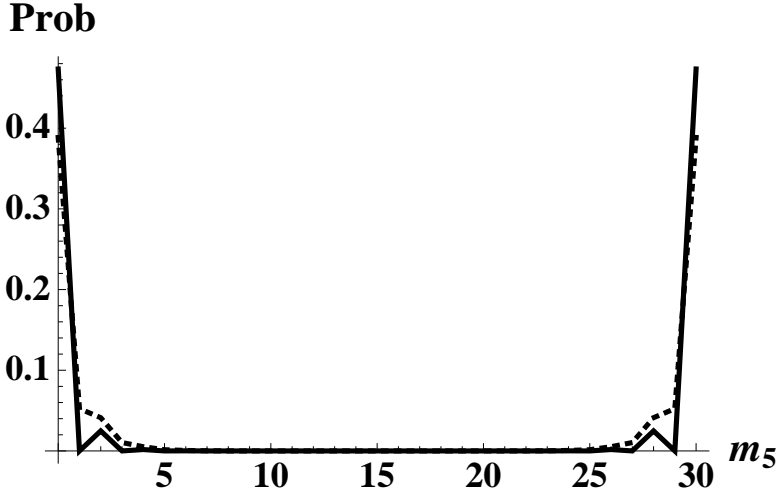}

\caption{Plot of $P_{m_{1},m_{2}m_{5},m_{6}}$ of Eq.
(\ref{eq:Prob56}) versus $m_{5}$ ($m_{6}=30-m_{5})$ for
$N_{\alpha}=N_{\beta}=30,$ $\xi=0$, and $m_{1}=m_{2}=15$ (solid line).
Also we show $m_{1}=18$, $m_{2}=12$ (dotted line) to show the relative
insensitivity of the NOON state to the values of $m_{1}$ and $m_{2}$.
A similar change is found for $N_{\alpha}=26$, $N_{\beta}=34$ with equal
$m_{1}$, $m_{2}$.  }

\label{figNOON}
\end{figure}

We next compute Eq. (\ref{C}) for the case $i=5$, $j=6.$ Instead
of introducing the phase state (\ref{eq:PhaseExp}) we expand the
binomial operator forms of Eq. (\ref{aops}) and compute the matrix elements involving
$a_{\alpha}$ and $a_{\beta}$, which involves appropriate $\delta$-functions;
we keep the quantity in the form of a sum resulting in the probability\begin{eqnarray}
P_{m_{1},m_{2}m_{5},m_{6}} & = & Km_{5}!m_{6}!\left|\sum_{p,q,r}\frac{e^{-i(p+q)(\xi-\pi/2)}(-1)^{p+r}}{p!(m_{1}-p)!q!(m_{2}-q)!r!(m_{5}-r)!}\right.\nonumber \\
 &  & \times\left.\frac{1}{(N_{\alpha}-p-q-r)!(p+q+r+m_{6}-N_{\alpha})!}\right|^{2}\label{eq:Prob56}\end{eqnarray}
By adjusting the phase $\xi$ to zero we get the NOON states as seen in Fig.
\ref{figNOON} for the case $m_{1}=m_{2}$ and for a case where this
optimal situation does not occur. 
In each case it is shown that the probability has small peaks away from 0 or 30,
but is very close to an ideal NOON state. Reference 9 proposes a "feedforward" method to make the two components of the phase superposition orthogonal, to produce a NOON state even when $m_{1}$ and $m_{2}$ are not equal.

\section{Probing the state}

\begin{figure}[h]
\centering
\includegraphics[width=4in]{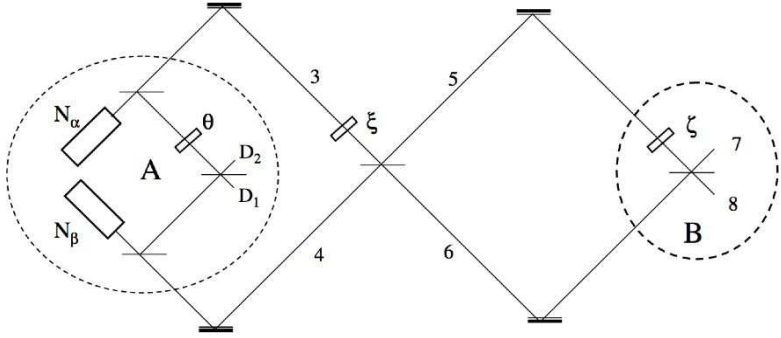}

\caption{The interferometer of Fig. \ref{fig1} extended to allow the
interference of the NOON states  at the beam splitter
for detectors 7 and 8. The resulting interference pattern is shown in
the figure below. }

\label{BigInterf}
\end{figure}

We now let the two states pass through arms 5 and 6 to interfere at a
final beam splitter and proceed to detectors 7 and 8 as shown in Fig.
\ref{BigInterf}. We now need the operators
\begin{eqnarray}
a_{7} & = & \frac{1}{2\sqrt{2}}\left(ue^{i\xi}a_{\alpha}+va_{\beta}\right)\quad a_{8}=\frac{1}{2\sqrt{2}}\left(ve^{i\xi}a_{\alpha}-ua_{\beta}\right)
\end{eqnarray}
where \begin{eqnarray}
u & = & \left(e^{i\zeta}-1\right)\nonumber \\
v & = & -i\left(e^{i\zeta}+1\right)
\end{eqnarray}

\begin{figure}[h]
\centering \includegraphics[width=4in]{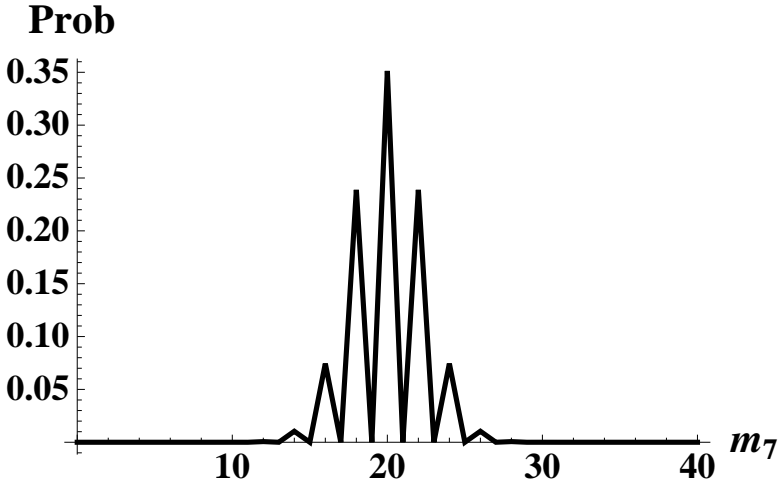}

\caption{Plot of $P_{m_{1},m_{2}m_{7},m_{8}}$ of Eq. (\ref{eq:P78uzero})
versus $m_{7}$ for $m_{1}=m_{2}=20$, $N_{\alpha}=N_{\beta}=40.$ Here
$\zeta=0$. 
}

\label{figPopOsc}
\end{figure}

We consider the phase angle $\zeta=0$ (we also have $\theta=\pi/2$ and $\xi=0$
as above) in which case we find\begin{equation}
P_{m_{1},m_{2},m_{7},m_{8}}=\frac{K}{m_{7}!m_{8}!}\left|\sum_{p=0}^{m_{1}}\frac{(-1)^{p}}{p!(m_{1}-p)!(N_{\alpha}-p-m_{8})!(m_{2}+m_{8}-N_{\alpha}+p)!}\right|^{2}\label{eq:P78uzero}\end{equation}
A plot of this probability versus $m_{7}$ is shown in Fig.
\ref{figPopOsc}.  The oscillations are equivalent to interference
fringes and are similar to those found in Ref. 7 where
the phase states shown
in Fig. \ref{figQ} were allowed to interfere. %

Leggett \cite{LeggettPRL} proposed that there were two basic assumptions
made implicitly by most physicists about the notion of reality at
the macroscopic level. 

A1) {}``Macroscopic realism: A macroscopic system with two or more
macroscopically distinct states available to it will at all times
be in one or the other of these two states.''

A2) {}``Noninvasive measurability at the macroscopic level: It is
possible, in principle, to determine the state of the system with
arbitrarily small perturbation on its subsequent dynamics.'' 

The quantum system we have studied can be used to test these rules of realism.
We can attempt to demonstrate experiments that might test these
assumptions using our macroscopic system.  We place beam splitters in
the arms 5 and 6 leading to side detectors $5^{\prime}$ and
$6^{\prime}$ (not shown in Fig.  \ref{BigInterf} but constructed much
like side detectors 1 and 2).  Since the probability is that perhaps
less than one particle might end in arm 5 while all the rest are in
the other arm, or vice versa, we might expect that, say, detecting
four particles in side detector $5^{\prime}$ (and none in detector
$6^{\prime}$) would ruin the interference pattern, and indeed it does.
The resulting plot has a single maximum with no oscillations.
Moreover, if we detect even just \emph{one} particle in detector
$5^{\prime}$ and none in $6^{\prime}$ the interference pattern is
ruined.  A non-invasive measurement process is not possible in the
quantum system; any probing disturbs the interference pattern, which
is a not-unexpected result of decoherence in quantum mechanics.  An
exception to this is if we find equal numbers of particles in
detectors $5^{\prime}$ and $6^{\prime}$; then the interference pattern
is re-established.  It is as if we just had a superposition of
symmetrically smaller states passing through the arms; we still do not
know via which arm the majority of the particles passed.

We next consider a negative experiment in which the beam splitter
leading to side detector $5^{\prime}$ has zero transmission probability
so that any particles in arm 5 are diverted into this side detector.
However, we consider \emph{only} situations in which $m_{5^{\prime}}=0,$
that is, no particles actually come into arm 5. According to A1, this
corresponds to a case in which all the particles have gone the other
way through arm 6. In at least half the experiments the blocking of
arm 5 should have no effect at all according to the realist proposition
A1. And yet in this case we get no interference pattern. Even though
no particles came into arm 5, and all the particles proceeded to the
beam splitter via arm 6, the interference is ruined,  in contrast to
assumption A1. Again this is not surprising in our quantum system.

\section{Conclusion}

Using double Bose condensates we have shown how to construct NOON
states, which are known to have useful applications, by state-vector
reduction and conditional preparation.  Bose condensate
interferometers, which we would require in our set-up, have already
been constructed.  \cite{BoseInterf} Other methods have been proposed
to create NOON states by projective measurement, with a number of
measurements that is an increasing function of the number of particles.
 \cite{Kok, Fiur}  Our method makes use of a single measurement and does not
even require precise values of $m_{1}$ and $m_{2}$, since the only requirement
is that they should be equal (or not very different, since the method
is relatively robust).  Here we have considered how NOON states might
be used to test experimentally the tenets of macroscopic realism and
to demonstrate macroscopic quantum interference.  In other work to be
reported elsewhere we have also applied these states to the
measurement of phase and have shown they can easily exceed the
classical limit and nearly reach the Heisenberg limit.
\cite{F&LHeisenberg}

\section*{Acknowledgement}

We thank Dr. Hugo Cable for pointing out the existence of Ref. 9 to us after we had submitted this paper. We also thank him for very useful discussions.

\end{document}